\documentclass[letterpaper]{article}

\usepackage[T1]{fontenc}

\usepackage{geometry}
\usepackage{setspace}

\usepackage{achemso}

\usepackage{graphicx}
\usepackage{float}
\newfloat{scheme}{htbp}{los}
\floatname{scheme}{Scheme}
\newfloat{graph}{htbp}{loh}
\usepackage{amsmath,amssymb}
\usepackage{placeins}

\usepackage[hidelinks]{hyperref}
\hypersetup{
  pdftitle={Grain Boundary Engineering Effect on Vortex Matter in Superconducting Films},
  pdfauthor={Qun Wang et al.}
}
\newcommand*\figref[2]{\hyperref[fig:#1]{Figure~\ref*{fig:#1}#2}}

\usepackage{authblk}
\author[1]{Qun Wang}
\author[1]{Ting Chen}
\author[1]{Ya-Xun He}
\author[1]{Xing-Jian Liu}
\author[1]{Jian-Wen Sun}
\author[1]{Kang-Hong Yin}
\author[2]{Fang-Ting Lin\textsuperscript{*}}
\author[1]{Shi-Xun Cao\textsuperscript{*}}
\author[1,3]{Jun-Yi Ge\textsuperscript{*}}
\affil[1]{Materials Genome Institute, Shanghai University, Shanghai 200444, China}
\affil[2]{Department of Physics, Shanghai Normal University, Shanghai 200234, China}
\affil[3]{Department of Physics and Shanghai Key Laboratory for High Temperature Superconductors, Shanghai University, Shanghai 200444, China}

\title{Grain Boundary Engineering Effect on Vortex Matter in Superconducting Films}
\date{\small \textsuperscript{*}Corresponding authors. E-mail: \href{mailto:nounou7@163.com}{nounou7@163.com} (Fang-Ting Lin); \href{mailto:sxcao@shu.edu.cn}{sxcao@shu.edu.cn} (Shi-Xun Cao); \href{mailto:junyi_ge@t.shu.edu.cn}{junyi\_ge@t.shu.edu.cn} (Jun-Yi Ge).}

\begin{document}
\maketitle

\begin{abstract}
Grain boundaries (GBs) in polycrystalline superconducting films act as a double-edged sword: they can pin vortices or degrade superconductivity through Josephson-like weak-link coupling. Here, we demonstrate that sputtering pressure tunes GB coupling in NbTiN films and visualize its consequences for vortex matter. The 5 mTorr film exhibits dispersed grain orientations and a two-step resistive transition under field, signaling intergranular weak-link behavior. In contrast, the 7 mTorr film develops a (111) texture, a single-step transition, higher critical current density, a second magnetization peak, and a \(\delta l\)-type pinning response consistent with improved GB coupling. Cryogenic magnetic force microscopy reveals a spatially heterogeneous, cluster-like vortex configuration in the 5 mTorr film, whereas the 7 mTorr film hosts a more uniform distribution with enhanced local order. These results establish a connection between deposition-controlled GB connectivity, macroscopic weak-link transport, and microscopic vortex organization, providing a route to tailor vortex pinning in polycrystalline superconducting films.
\end{abstract}

\section*{Keywords}

NbTiN thin films; grain-boundary connectivity; weak-link effect; vortex pinning; magnetic force microscopy

\section{Introduction}

Grain boundaries (GBs) are among the most decisive microstructural
elements controlling the macroscopic performance of superconducting
films. In high-\emph{T}\textsubscript{c} cuprates and Fe-based
superconductors, even a single misoriented GB can act as a Josephson
weak link and suppress the intergranular critical current by orders of
magnitude.\cite{ref1,ref2,ref3,ref4} In conventional s-wave superconductors,
the suppression is generally weaker, yet GBs still strongly affect
normal-state scattering, the vortex-pinning landscape, and the spatial
homogeneity of the order parameter.\cite{ref5,ref6,ref7} Engineering GB
networks therefore offers a route to tailor superconducting transport
without changing the chemistry of the host material---a particularly
attractive strategy for thin-film-based superconducting electronics and
single-photon detectors based on NbTiN.\cite{ref8,ref9} Despite
extensive transport studies,\cite{ref10,ref11,ref12} a central question
remains open: how does continuous tuning of GB connectivity reorganize
vortex matter in real space? Macroscopic measurements such as
\emph{R}(\emph{T}), \emph{H}\textsubscript{c2}(\emph{T}), and
\emph{J}\textsubscript{c}(\emph{H}) integrate over the entire film and
cannot distinguish whether a change in pinning originates from
GB-localized vortex trapping, bulk pinning inside grains, or the
redistribution of supercurrents around weakly coupled grain clusters.
Direct real-space imaging of vortices in films with deliberately varied
GB connectivity---but otherwise identical composition, thickness, and
substrate---has remained scarce, leaving the microscopic correspondence
between GB coupling and vortex configuration largely indirect.

Here we address this gap using polycrystalline NbTiN films in which the effective GB coupling state is tuned through a single deposition parameter, the sputtering pressure. By varying the sputtering pressure between 5 and 7 mTorr while keeping all other growth conditions fixed, we obtain two films with nearly identical thickness and composition but distinct textures and effective GB coupling states: a 5 mTorr film with dispersed grain orientations and weaker intergranular coupling, and a 7 mTorr film with a pronounced (111) texture and improved intergranular coupling. This single-parameter comparison helps isolate the role of effective GB connectivity while minimizing variations in nominal chemistry and geometry. Transport measurements reveal a clear two-step superconducting transition under magnetic field in the weakly coupled film, consistent with intergranular Josephson decoupling, whereas the connectivity-improved film exhibits a single sharp transition. Critically, magnetic force microscopy (MFM) directly visualizes the corresponding vortex configurations: the 5 mTorr film hosts a more heterogeneous, cluster-like vortex arrangement, whereas the 7 mTorr film supports a more uniform distribution with enhanced local order. The two regimes therefore differ not only quantitatively in \(T_c\), \(H_{c2}\), and \(J_c\), but also qualitatively in the geometry of the vortex state.

These results establish pressure-controlled effective GB connectivity as a practical tuning knob for the vortex-pinning response of a conventional s-wave superconductor and provide a direct real-space correspondence between intergranular weak-link behavior and vortex configuration. Beyond the specific case of NbTiN, the methodology demonstrates that real-space vortex imaging combined with single-parameter GB tuning can help resolve transport-level ambiguities concerning the microscopic origin of pinning, with implications for superconducting thin films used in single-photon detection, kinetic-inductance devices, and superconducting qubits.

\section{Experimental Methods}

The NbTiN thin films were fabricated by reactive magnetron sputtering on
silicon substrates using a high-vacuum system (FSE-CLS-SP-100, Fulin
Technology Co., Ltd.). The base pressure was 5 $\times$ 10\textsuperscript{$-$7} Torr. A Nb:Ti
alloy target (70:30 wt\%) was sputtered in a mixture of high-purity Ar
(99.999\%) and N\textsubscript{2} (99.999\%). The Ar and N\textsubscript{2} flow rates were maintained
at 81 and 9 sccm, respectively. The target--substrate distance was fixed
at 70 mm, and the substrates were maintained at room temperature without
intentional heating. Two films were deposited at total sputtering
pressures of 5 and 7 mTorr. Their thicknesses were controlled at
approximately 160 nm by adjusting the deposition time. The film
thicknesses were measured by stylus profilometry. All deposition
conditions other than the total sputtering pressure were kept fixed.

Crystal structure was characterized by X-ray diffraction (XRD).
Electrical and magnetic properties were measured using a Physical
Property Measurement System (PPMS-14T, Quantum Design) and a Magnetic
Property Measurement System (MPMS-7T, Quantum Design), respectively.
Vortex imaging was performed with a cryogenic magnetic force microscope
(attocube). Before imaging, each sample was field cooled from above
\(T_{c}\) to 1.7 K. Magnetic contrast was acquired in noncontact lift
mode through the cantilever phase signal, following established
low-temperature MFM approaches for superconducting
vortices.\cite{ref13,ref14} In all magnetic measurements, the
applied field was perpendicular to the film surface.

\section{Results and Discussion}

\begin{figure}[!htbp]
  \centering
  \includegraphics[width=0.92\linewidth]{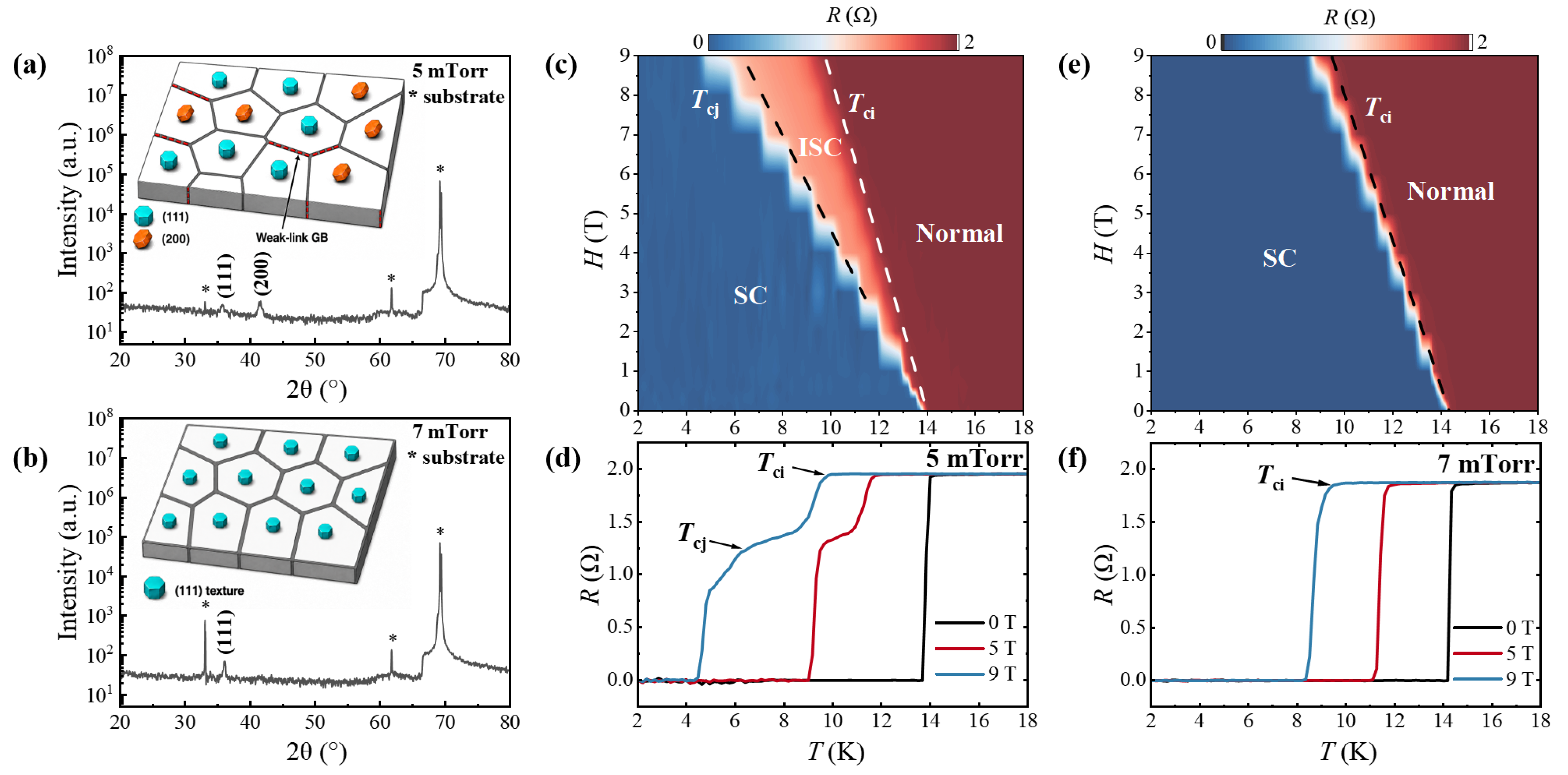}
  \caption[Figure 1]{\textbf{Structural and transport
characterization of NbTiN films with different grain-orientation
distributions and GB connectivity.} \textbf{(a,b)} XRD patterns of the 5
and 7 mTorr films, respectively. The 5 mTorr film shows comparable
(111) and (200) reflections, whereas the 7 mTorr film exhibits a
pronounced (111) texture. These texture differences are associated with
the distinct effective GB coupling states identified by transport.
Insets schematically illustrate the orientation distributions and
corresponding GB networks.
\textbf{(c,e)} \(H\)--\(T\) resistance maps of the 5 and 7 mTorr films.
Dashed lines mark the characteristic transition boundaries; \(T_{ci}\)
and \(T_{cj}\) are distinguished for the 5 mTorr film, corresponding to
the superconducting transitions of intragranular regions and the
grain-boundary network, respectively. The 7 mTorr film only shows one
dominant boundary. ISC denotes the intermediate isolated-superconducting
regime. \textbf{(d,f)} Representative \(R\)--\(T\) curves measured at 0,
5, and 9 T for the 5 and 7 mTorr films, respectively.}
  \label{fig:1}
\end{figure}

The NbTiN films deposited at 5 and 7 mTorr exhibit distinct out-of-plane
crystallographic textures, as revealed by the XRD patterns in
\figref{1}{(a,b)}. Both films display the
characteristic diffraction peaks of NbTiN, but their preferred
orientations differ markedly. For the 5 mTorr film, the (111) and (200)
peak intensities are comparable, indicating a relatively dispersed
orientation distribution without a dominant out-of-plane texture. In
contrast, the (111) reflection dominates the pattern of the 7 mTorr
film, whereas the (200) reflection is strongly suppressed and approaches
the background level, indicating the development of a pronounced (111)
texture. Thus, under the present deposition conditions, increasing the
total sputtering pressure from 5 to 7 mTorr favors (111)-oriented
growth.

The texture difference is consistent with different statistical distributions of local grain misorientation and intergranular coupling. Because conventional \(\theta\)--\(2\theta\) XRD does not resolve individual GB angles, the insets are conceptual rather than direct GB maps; the effective coupling regimes are established primarily from field-dependent transport.

The normalized \(R\)--\(T\) curves measured at 0, 5, and 9 T reveal
distinct field-dependent transitions {[}\figref{1}{(d,f)}{]}. The normal-state resistance \(R_{n}\) was taken immediately
above the superconducting transition, and the onset temperature
\(T_{ci}\) was defined by the \(R/R_{n} = 0.90\) criterion. At zero
field, both films exhibit sharp transitions, with
\(T_{ci} \approx 13.96\) K for the 5 mTorr film and
\(T_{ci} \approx 14.13\) K for the 7 mTorr film.

With increasing field, the 5 mTorr film develops a two-step transition,
a well-established transport signature of granular or weakly coupled
superconductors.\cite{ref4,ref15} We denote the higher-temperature
onset by \emph{T}\textsubscript{ci}, corresponding to the establishment
of intragranular or local superconductivity, and the lower-temperature
shoulder by \emph{T}\textsubscript{cj}, associated with the gradual
establishment of an intergranular Josephson-coupled current
path.\cite{ref1,ref3} Because this lower-temperature feature is a
broadened shoulder rather than a sharp transition,
\emph{T}\textsubscript{cj} is used as a qualitative marker and is not
assigned a fixed resistance criterion. Its more rapid shift to lower
temperature relative to \emph{T}\textsubscript{ci} indicates that
intergranular coupling is more sensitive to the applied field.

In contrast, the 7 mTorr film maintains a predominantly single-step
superconducting transition over the measured field range. The transition
shifts to lower temperature with increasing field without a clearly
resolved low-temperature shoulder. This behavior indicates that the
weak-link contribution is substantially reduced, producing more uniform
intergranular coupling and improved global superconducting connectivity.
The absence of a resolvable \emph{T}\textsubscript{cj} feature does not
imply that all GB effects are eliminated; rather, they no longer
generate a distinct second resistive transition within the experimental
resolution.

The complete \(R\)--\(T\) data from 0 to 9 T were used to construct the
resistive \(H\)--\(T\) maps in \figref{1}{(c,e)}.
For the 5 mTorr film, an intermediate resistive plateau separates
\(T_{ci}\) and \(T_{cj}\). This isolated-superconducting (ISC) regime is
interpreted as a state in which local superconductivity has formed but
intergranular coupling is still insufficient to establish a globally
coherent path. The 7 mTorr map does not display a resolvable \(T_{cj}\)
boundary and is therefore divided into normal and superconducting
regimes.

The upper critical field was estimated from the onset line, with
\(T_{c} = T_{ci}\), using the dirty-limit Werthamer--Helfand--Hohenberg
(WHH) orbital expression\cite{ref16}:
\(\mu_{0}H_{c2}(0) = - 0.693\, T_{c}\left. \frac{d\left( \mu_{0}H_{c2} \right)}{dT} \right|_{T = T_{c}}\).

The zero-temperature coherence length was then calculated
from\cite{ref17}:
\(\xi(0) = \left[ \frac{\Phi_{0}}{2\pi\mu_{0}H_{c2}(0)} \right]^{1/2}\),
where \(\Phi_{0}\) is the flux quantum. The resulting values are
\(\mu_{0}H_{c2}(0) = 9.13\) T and 16 T for the 5 and 7 mTorr films,
corresponding to \(\xi(0) = 6.0\) and 4.53 nm, respectively. These
differences indicate distinct effective superconducting length scales
and field-suppression rates, plausibly reflecting pressure-induced
changes in microstructure and scattering. The low-temperature \(T_{cj}\)
shoulder is used only as an indicator of intergranular coupling and is
not included in the \(H_{c2}\) estimate.

\begin{figure}[!htbp]
  \centering
  \includegraphics[width=0.92\linewidth]{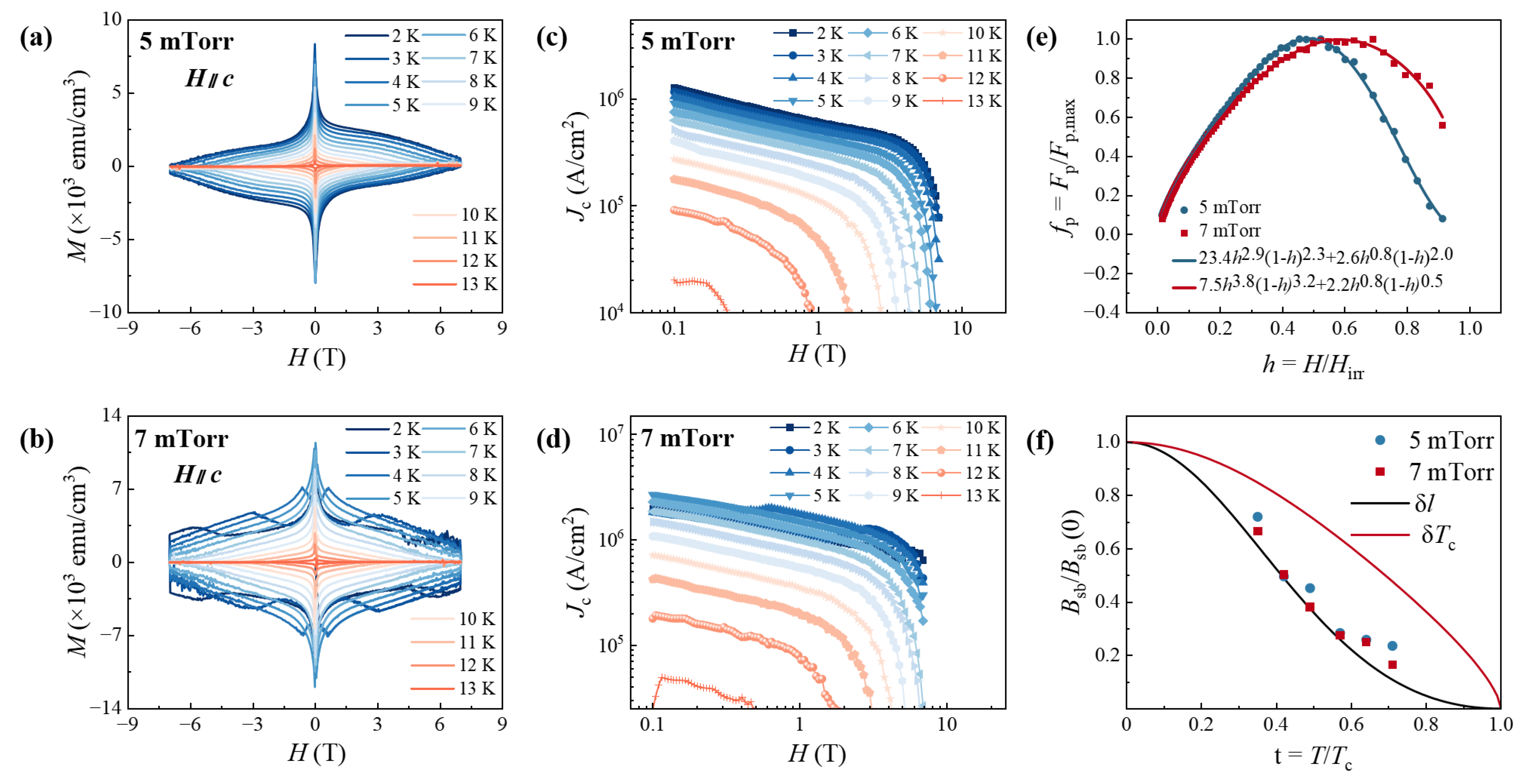}
  \caption[Figure 2]{\textbf{Magnetic properties and
vortex-pinning response of the NbTiN films.} \textbf{(a,b)} Isothermal
\(M\)--\(H\) loops of the 5 and 7 mTorr films. \textbf{(c,d)} Field
dependence of \(J_{c}\), calculated from the hysteresis loops using the
extended Bean model. \textbf{(e)} Normalized pinning force
\(f_{p} = F_{p}/F_{p,max}\) as a function of reduced field
\(h = H/H_{irr}\); solid curves are two-component Dew--Hughes fits.
\textbf{(f)} Temperature dependence of the single-vortex-to-small-bundle
crossover field \(B_{sb}\); solid curves are fits to the
collective-pinning expression.}
  \label{fig:2}
\end{figure}

The isothermal magnetization loops in \figref{2}{(a,b)} reveal a stronger irreversible response in the 7 mTorr film,
whose loops are wider and develop a second magnetization peak (SMP) at
intermediate-to-high fields. SMP or fishtail features are generally
associated with a field-driven change in vortex pinning, elasticity, or
creep regime, although their detailed microscopic origin is material
dependent.\cite{ref18,ref19,ref20} In the present films, the SMP appears
together with improved global connectivity, allowing the irreversible
current response to persist to higher fields. The data alone do not
establish a unique elastic-to-plastic crossover, because
magnetic-relaxation measurements would be required for that assignment.

The 5 mTorr film shows a weaker irreversible response and no comparably pronounced SMP, consistent with weak-link-limited global current transport. These correlations do not identify a unique microscopic origin of the SMP.

The critical current density was obtained from the hysteresis width
using the extended Bean expression for a rectangular
specimen,\cite{ref21,ref22}
\(J_{c} = \frac{20\,\Delta M}{a(1 - a/3b)},\ a < b\),
where \emph{\ensuremath{\Delta}M} is in emu cm\textsuperscript{$-$3} and the in-plane
dimensions \emph{a} and \emph{b} are in cm, giving
\emph{J}\textsubscript{c} in A cm\textsuperscript{$-$2}. The 7 mTorr film
exhibits the larger \emph{J}\textsubscript{c} throughout the measured
field range {[}\figref{2}{(c,d)}{]}, while
\emph{J}\textsubscript{c} decreases with increasing temperature in both
films. The higher \emph{J}\textsubscript{c} of the 7 mTorr film is
consistent with a reduction of weak-link-limited current paths and
improved macroscopic current connectivity.

The pinning-force density was calculated as
\(F_{p} = J_{c}B \approx \mu_{0}HJ_{c}\) and normalized by its maximum
value. \figref{2}{(e)} plots \(f_{p}\) against
\(h = H/H_{irr}\), where \(H_{irr}\) was obtained by extrapolating the
high-field portion of \(J_{c}(H)\) to the criterion \(J_{c} = 100\) A
cm\textsuperscript{$-$2}. In the Dew--Hughes framework, a single
contribution is written\cite{ref23,ref24} as
\(f_{p} = Ah^{p}(1 - h)^{q}\), with \(h_{0} = p/(p + q)\).

Within the Dew--Hughes normal-core classification, \(h_{0}\approx 0.20\) and 0.33 are associated with surface/planar and point-like pins, whereas higher values may reflect higher-field volume or \(\Delta\kappa\)-type contributions. The peak position alone does not uniquely distinguish \(\delta l\) from \(\delta T_{c}\) pinning.

Both films exhibit broad \(f_{p}(h)\) curves that are not described over
the full field range by one component. We therefore use the empirical
two-component form
\(f_{p} = A_{1}h^{p_{1}}(1 - h)^{q_{1}} + A_{2}h^{p_{2}}(1 - h)^{q_{2}}\).

For the 5 mTorr film, the fit gives \emph{A}\textsubscript{1} = 23.4,
\emph{p}\textsubscript{1} = 2.9, \emph{q}\textsubscript{1} = 2.3,
corresponding to \emph{h}\textsubscript{01} $\approx$ 0.56, together with
\emph{A}\textsubscript{2} = 2.6, \emph{p}\textsubscript{2} = 0.8,
\emph{q}\textsubscript{2} = 2.0. The second component
(\emph{h}\textsubscript{02} $\approx$ 0.29) is compatible with a lower-field
point-like normal-core contribution, whereas the dominant component is
shifted to higher reduced field. For the 7 mTorr film, the fit gives
\emph{A}\textsubscript{1} = 7.5, \emph{p}\textsubscript{1} = 3.8,
\emph{q}\textsubscript{1} = 3.2, corresponding to
\emph{h}\textsubscript{01} $\approx$ 0.54, together with
\emph{A}\textsubscript{2} = 2.2, \emph{p}\textsubscript{2} = 0.8,
\emph{q}\textsubscript{2} = 0.5. The associated
\emph{h}\textsubscript{02} $\approx$ 0.62 component is consistent with the
high-field shift of the pinning-force maximum and the SMP. These
empirical components describe the broad field dependence but should not
be assigned to unique microscopic defects solely from their peak
positions.

The crossover from single-vortex to small-bundle pinning is
characterized by \emph{B}\textsubscript{sb}(\emph{T})
{[}\figref{2}{(f)}{]}. The crossover was
identified from the change in the field dependence of
$-$ln{[}\emph{J}\textsubscript{c}(\emph{B})/\emph{J}\textsubscript{c}(0){]}
used to distinguish the single-vortex and small-bundle regimes.

Within collective-pinning theory\cite{ref25,ref26}:
\(B_{sb}(T) = B_{sb}(0)\left( \frac{1 - t^{2}}{1 + t^{2}} \right)^{\nu}\),
where \(t = T/T_{c}\), and \emph{\ensuremath{\nu}} = 2 and 2/3 are the limiting forms commonly used
for \emph{\ensuremath{\delta}l}- and \emph{\ensuremath{\delta}T}\textsubscript{c}-type pinning,
respectively. The fitted values are \emph{\ensuremath{\nu}} $\approx$ 1.5 for the 5 mTorr film
and \emph{\ensuremath{\nu}} $\approx$ 1.8 for the 7 mTorr film. Both are closer to the
\emph{\ensuremath{\delta}l} limit, indicating that mean-free-path fluctuations associated
with structural disorder make the larger contribution, whereas their
deviation from 2 indicates a mixed pinning response.\cite{ref27}
The value for the 7 mTorr film is closer to the \emph{\ensuremath{\delta}l} limit,
consistent with a more coherent high-field current response after
suppression of the weak-link network. This comparison identifies the
dominant fluctuation type within the adopted collective-pinning model;
it does not assign a unique microscopic defect species.

Low-temperature MFM was used to image field-cooled vortex configurations
and thereby probe the spatial consequences of the two coupling regimes.
Vortex positions respond to both vortex--vortex repulsion and the local
pinning potential, making real-space imaging complementary to the
macroscopic \(J_{c}\) analysis.\cite{ref6,ref28,ref29}

\begin{figure}[!htbp]
  \centering
  \includegraphics[width=0.88\linewidth]{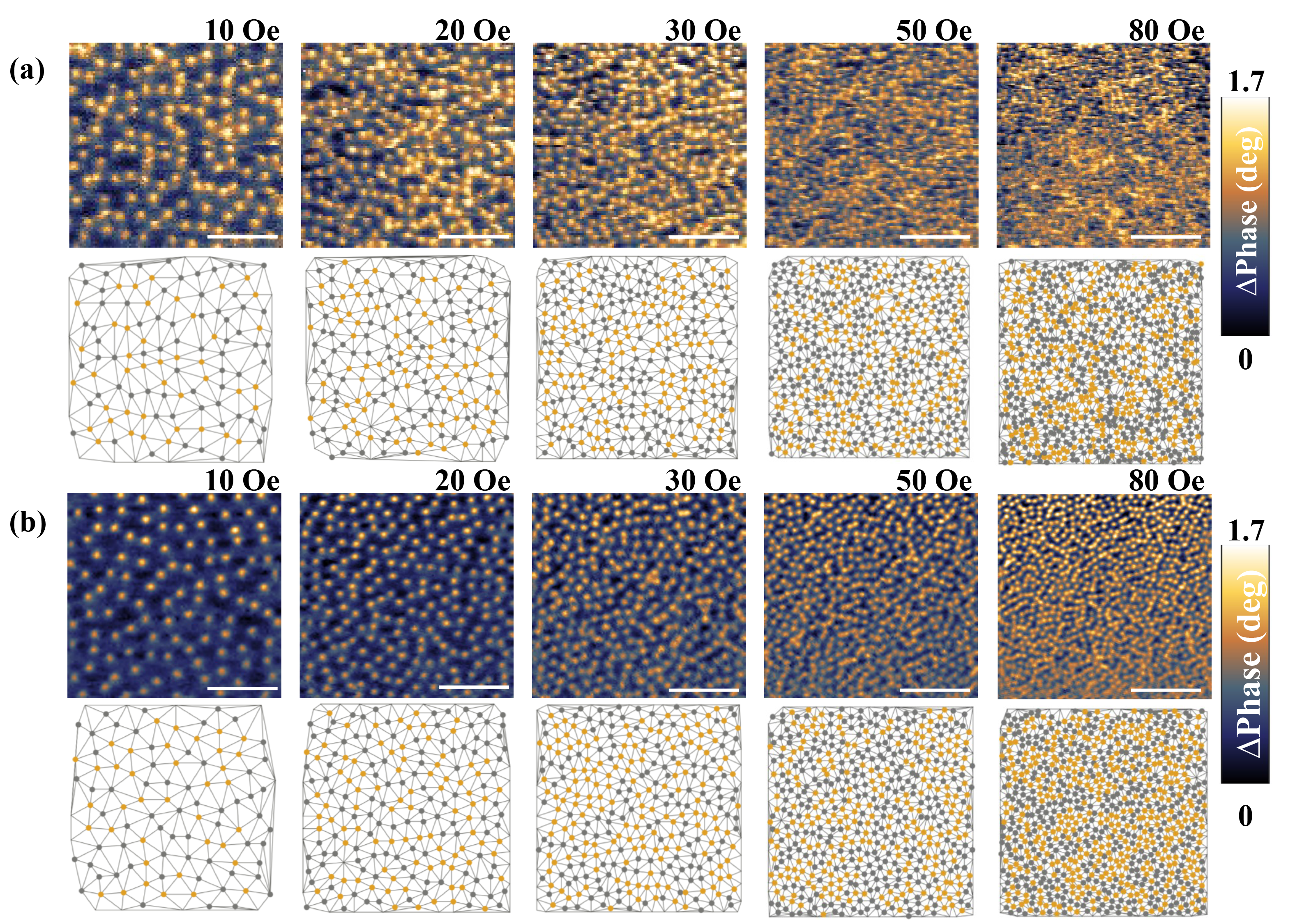}
  \caption[Figure 3]{\textbf{Real-space vortex
configurations and corresponding Delaunay triangulations.} MFM phase
images and triangulations for (a) the 5 mTorr and (b) the 7 mTorr films.
Bright features represent individual vortices according to the selected
MFM phase contrast. Images were acquired at 1.7 K after field cooling in
10--80 Oe. For each field, the Delaunay triangulation obtained from the
extracted vortex positions is shown below the MFM image. Yellow circles
denote sixfold-coordinated vortices and gray circles denote other
coordination numbers. Scale bars: 5 \(\text{\ensuremath{\mu}}\)m.}
  \label{fig:3}
\end{figure}

\figref{3}{} shows MFM phase images acquired at
1.7 K after field cooling in \emph{H}\textsubscript{FC} = 10--80 Oe,
together with Delaunay triangulations of the extracted vortex
coordinates.\cite{ref30,ref31} Sixfold- and
non-sixfold-coordinated vortices are marked by yellow and gray points,
respectively. In the 5 mTorr film, individual vortices are readily
identifiable at low fields, but their spatial distribution is visibly
heterogeneous. As the field increases, local crowding and sparse regions
make vortex identification increasingly difficult. In contrast, the 7
mTorr film exhibits a more homogeneous distribution over the same field
range, without obvious large-scale clustering or comparably sparse
regions. This contrast is consistent with suppression of a spatially
nonuniform, weak-link-related pinning landscape when the GB connectivity
is improved.

To quantify these differences, we evaluated nearest-neighbor distances,
Delaunay bond angles and coordination, the pair-correlation function
\(g(r)\), and the local sixfold bond-orientational order parameter
\(\psi_{6}\). These metrics are widely used to distinguish positional
and orientational order in two-dimensional vortex
configurations.\cite{ref31,ref32,ref33,ref34}

\begin{figure}[!htbp]
  \centering
  \includegraphics[width=0.88\linewidth]{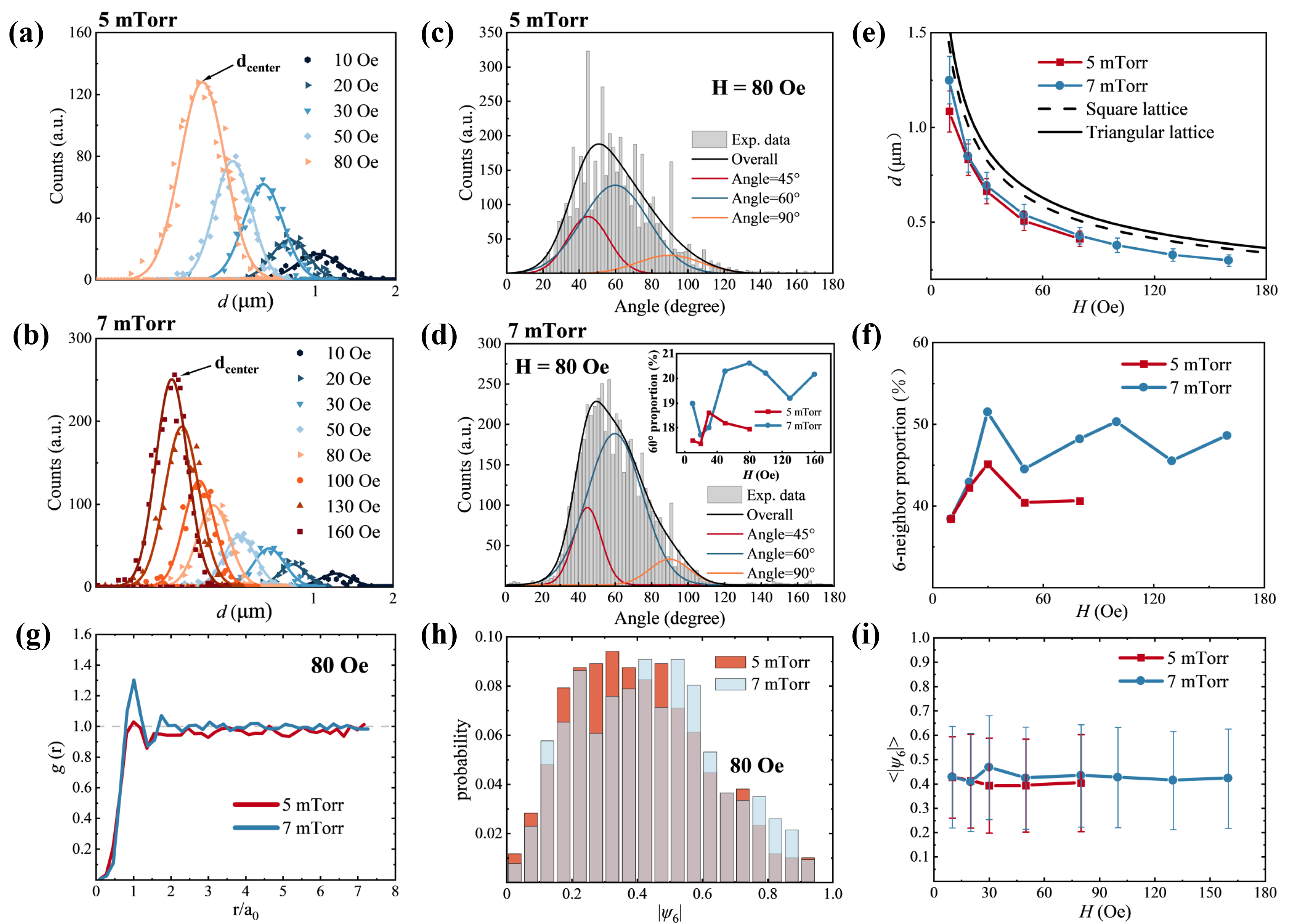}
  \caption[Figure 4]{\textbf{Quantitative comparison
of vortex ordering.} (a,b)
Nearest-neighbor distance distributions at different fields for the 5
and 7 mTorr films; solid curves are Gaussian fits. (c,d) Delaunay
bond-angle distributions at 80 Oe. Colored curves are Gaussian
components centered at 45°, 60°, and 90°, and the dashed curve is their
sum. (e) Field dependence of d\textsubscript{center}, compared with the
theoretical intervortex spacings for ideal square and triangular
lattices. Bars indicate the fitted distribution widths at 90\% of the
corresponding peak maxima and are not statistical uncertainties. (f)
Fraction of sixfold-coordinated vortices. (g) Pair-correlation functions
g(r) at 80 Oe, plotted against r/a\textsubscript{0}; the horizontal
dashed line marks g(r) = 1. (h) Histograms of
\textbar \ensuremath{\psi}\textsubscript{6}\textbar{} at 80 Oe. (i) Field dependence of
\textless\textbar \ensuremath{\psi}\textsubscript{6}\textbar\textgreater; error bars are
the standard deviations within each analyzed vortex ensemble.}
  \label{fig:4}
\end{figure}

The nearest-neighbor distributions in \figref{4}{(a,b)} shift to smaller distances as \(H_{\mathrm{FC}}\) increases, and their fitted peak positions define \(d_{\mathrm{center}}\). Over the common 10--80 Oe range, the widths are comparable and do not establish an ordering difference by themselves. The 7 mTorr film remains analyzable to 160 Oe and generally has a slightly larger \(d_{\mathrm{center}}\) at the same field, consistent with weaker local clustering within the analyzed field of view; this relative descriptor is not used alone to identify the pinning mechanism.

The bond-angle distributions at 80 Oe are compared in
\figref{4}{(c,d)}. An ideal triangular vortex
lattice has a characteristic nearest-neighbor angle of
60°.\cite{ref35} The 45° and 90° components are used here to
parameterize non-60° local arrangements rather than to assign unique
lattice symmetries. Both films contain multiple components and therefore
lack ideal long-range triangular order. The 7 mTorr film exhibits a more
pronounced 60° contribution and a higher 60° fraction at most fields,
indicating improved local triangular ordering. Individual-field
fluctuations are expected from the finite vortex number, local pinning
variations, and boundary sensitivity of the triangulation and should not
be interpreted as independent field-dependent transitions.

\noindent\parbox{\linewidth}{%
\hspace*{\parindent}\figref{4}{(e)} compares
\emph{d}\textsubscript{center} with the ideal square-lattice spacing
\emph{a}\textsubscript{sq} =
(\emph{\ensuremath{\Phi}}\textsubscript{0}/\emph{B})\textsuperscript{1/2} and the ideal
triangular-lattice spacing. The latter is
\(a_{0} = \left( \frac{2\Phi_{0}}{\sqrt{3}B} \right)^{1/2}\).
}

\noindent\parbox{\linewidth}{%
\hspace*{\parindent}For both films, \emph{d}\textsubscript{center} decreases with increasing
field but does not fully coincide with either ideal-lattice reference,
confirming the influence of disorder and pinning. The 7 mTorr values
generally lie closer to the triangular-lattice curve. The
sixfold-coordinated fraction {[}\figref{4}{(f)}{]}
is approximately 40\% for the 5 mTorr film and approximately 50\% for
the 7 mTorr film over most fields. These values indicate enhanced local
order in the 7 mTorr film but do not imply a defect-free, long-range
vortex lattice.
}

The pair-correlation functions at 80 Oe
{[}\figref{4}{(g)}{]} exhibit a first-neighbor
peak followed by rapidly damped oscillations, demonstrating short-range
positional correlation rather than long-range translational order. The
first-neighbor feature is slightly sharper for the 7 mTorr film. The
local sixfold order parameter was calculated as
\(\psi_{6}(i) = \frac{1}{N_{i}}\sum_{j = 1}^{N_{i}}\exp\left( 6i\theta_{ij} \right)\),
where \(N_{i}\) is the number of selected neighbors and
\(\theta_{ij}\) is the bond angle relative to a fixed axis. The broad
\(\left| \psi_{6} \right|\) histograms in
\figref{4}{(h)} show spatially nonuniform
orientational order in both films, while the 7 mTorr film has slightly
greater weight at larger \(\left| \psi_{6} \right|\). Its mean value
\(\langle\left| \psi_{6} \right|\rangle\) is also modestly higher at
most fields {[}\figref{4}{(i)}{]}. Taken together,
the distance, angular, coordination, pair-correlation, and orientational
analyses consistently support a more homogeneous and locally ordered
vortex configuration after improving the GB connectivity.

\noindent\parbox{\linewidth}{%
\hspace*{\parindent}The extracted vortex coordinates were further used to estimate local
vortex--vortex interaction energies and the corresponding force
imbalance. The 160 nm film thickness is comparable to, or smaller than,
the penetration depth expected near the effective vortex-freezing
temperature. A rigorous thin-film treatment would therefore use the
Pearl interaction, whose characteristic screening length is of order
\(\lambda^{2}/d\) (often written \(2\lambda^{2}/d\) depending on
convention).\cite{ref36,ref37} Because this interaction is
long-ranged, vortices outside the finite MFM field of view can make a
substantial and position-dependent contribution, particularly near the
image boundary.\cite{ref37,ref38} The values below are
consequently treated as semi-quantitative local descriptors rather than
absolute pinning energies or forces.
}

As an additional dimensionality check, the crossover thickness was
estimated from \(d_{c} = 2L_{c}\), with\cite{ref39,ref40}
\(L_{c} \approx \left[ \frac{2d\xi}{3\sqrt{3}\ln(d/\xi)} \right]^{1/2}\).

Taking \emph{\ensuremath{\xi}} = \emph{\ensuremath{\xi}}(0) for this estimate and using \emph{d} = 160
nm together with the zero-temperature coherence lengths calculated above
gives \emph{d}\textsubscript{c} $\approx$ 21 nm for the 5 mTorr film and
\emph{d}\textsubscript{c} $\approx$ 18 nm for the 7 mTorr film. Both values are
well below the actual film thickness, suggesting that the films are not
in the thickness-driven two-dimensional collective-pinning limit. We
therefore use the London vortex-line expressions as local near-neighbor
approximations, while retaining the caveat that the full electromagnetic
interaction is thin-film-like.

Within this approximation, the interaction energy per unit length
assigned to vortex \(i\) is\cite{ref17,ref41,ref42}
\(\varepsilon_{\mathrm{int},i} = \sum_{j \neq i}2\varepsilon_{0}K_{0}\left( \frac{r_{ij}}{\lambda} \right)\),
where \(\varepsilon_{0} = \Phi_{0}^{2}/(4\pi\mu_{0}\lambda^{2})\),
\emph{K}\textsubscript{0} is the zeroth-order modified Bessel
function and \emph{r}\textsubscript{ij} =
\textbar{}\emph{r}\textsubscript{i} $-$ \emph{r}\textsubscript{j}\textbar.
The corresponding net vortex--vortex interaction force per unit length
is
\(\mathbf{f}_{i} = \sum_{j \neq i}\frac{\Phi_{0}^{2}}{2\pi\mu_{0}\lambda^{3}}\frac{\mathbf{r}_{ij}}{r_{ij}}K_{1}\left( \frac{r_{ij}}{\lambda} \right)\).

In a mechanically metastable frozen configuration, the net
vortex--vortex interaction force must be balanced by pinning and other
forces. Accordingly, \textbar{}\emph{f}\textsubscript{i}\textbar{} is
used here as a comparative force-imbalance descriptor of local spatial
heterogeneity, rather than as a direct measurement of the absolute
pinning force.\cite{ref14}

\begin{figure}[!htbp]
  \centering
  \includegraphics[width=0.92\linewidth]{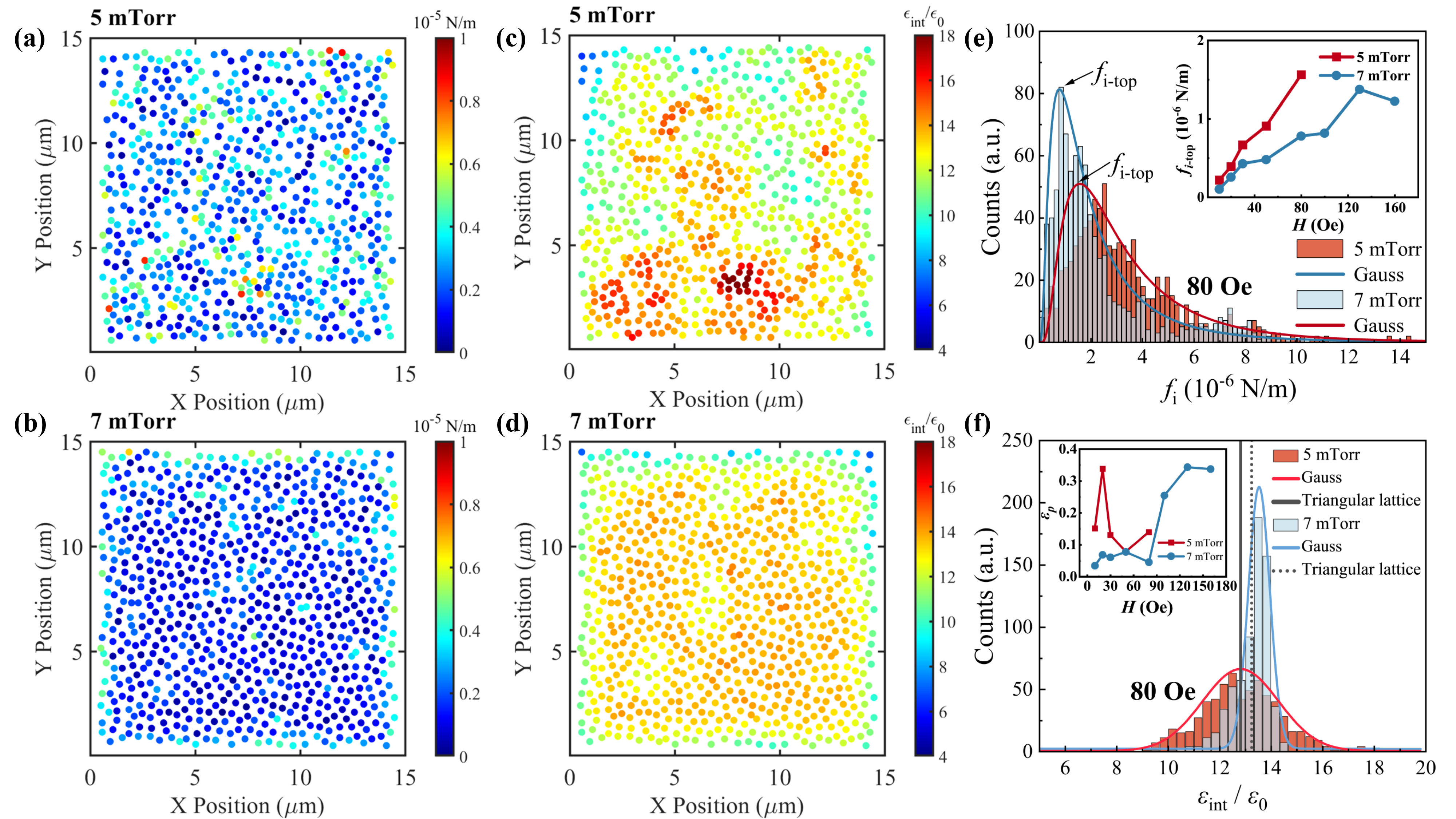}
  \caption[Figure 5]{\textbf{Local force imbalance
and vortex--vortex interaction energy derived from the measured vortex
coordinates. (a,b)} Spatial maps of the local force-imbalance magnitude
per unit vortex length,~\textbar f\textsubscript{i}\textbar, for the 5
and 7 mTorr films, respectively, at 80 Oe. \textbf{(c,d)} Corresponding
maps of the normalized local vortex--vortex interaction
energy,~\ensuremath{\varepsilon}\textsubscript{int}/\ensuremath{\varepsilon}\textsubscript{0},
where~\ensuremath{\varepsilon}\textsubscript{0}= \ensuremath{\Phi}\textsubscript{0}\textsuperscript{2}/(4\ensuremath{\pi}\ensuremath{\mu}\textsubscript{0}\ensuremath{\lambda}\textsuperscript{2}).
Identical color scales are used within each pair of maps to enable
direct comparison. \textbf{(e)} Statistical distributions
of~\textbar f\textsubscript{i}\textbar; solid curves show the empirical
fits used to determine the distribution centers. The inset summarizes
the field dependence of the fitted center,~\ensuremath{\mu}\textsubscript{f} .
\textbf{(f)} Statistical distributions
of~\ensuremath{\varepsilon}\textsubscript{int}/\ensuremath{\varepsilon}\textsubscript{0}. The black solid and dashed
vertical lines indicate the interaction energies calculated for ideal
triangular vortex lattices with the same vortex densities as those in
the 5 and 7 mTorr films, respectively. The inset shows the field
dependence of the deviation from the corresponding ideal-lattice
value,~\ensuremath{\Delta}\ensuremath{\varepsilon}\textsubscript{int}/\ensuremath{\varepsilon}\textsubscript{0} . Values were calculated
using the London vortex-line approximation and are used for
semi-quantitative comparison between films with similar thicknesses.
Only vortices located outside the boundary-exclusion region were
included in the statistical analysis.}
  \label{fig:5}
\end{figure}

To retain the numerical basis used for \figref{5}{}, we adopt \emph{\ensuremath{\lambda}}(0) = 310 nm as a literature-guided representative
model value for disordered NbTiN films, rather than as an independently
measured material constant of either sample.\cite{ref43} The
penetration depth at the effective freezing temperature was estimated
using the two-fluid expression\cite{ref17},
\(\lambda(T_{f}) = \frac{\lambda(0)}{\sqrt{1 - (T_{f}/T_{c})^{4}}}\).

An effective ratio \emph{T}\textsubscript{f}/\emph{T}\textsubscript{c} =
0.92 was selected as a representative calculation input. Together with
\emph{\ensuremath{\lambda}}(0) = 310 nm, this ratio is not treated as a measured material
constant for either film. Because neither \emph{T}\textsubscript{f} nor
\emph{\ensuremath{\lambda}}(0) was independently measured for these two films, the
resulting maps are used only for relative comparison under identical
assumptions. Sensitivity to
\emph{T}\textsubscript{f}/\emph{T}\textsubscript{c} and \emph{\ensuremath{\lambda}}(0) is
examined in Figures S1 and S2 and Table S2 of the Supporting
Information. Although the absolute energy and force-imbalance scales
vary with the assumed parameters, the broader interaction-energy
distribution of the 5 mTorr film remains unchanged over the tested
ranges.

The force-imbalance maps in \figref{5}{(a,b)} show
stronger spatial fluctuations in the 5 mTorr film, including localized
high-value regions adjacent to substantially weaker areas. The 7 mTorr
map is more spatially uniform. In conjunction with the two-step
transport transition, the greater heterogeneity of the 5 mTorr film is
consistent with a nonuniform GB-related pinning landscape. It is not,
however, a direct map of individual GBs because the structural and
vortex images were not spatially registered. Prior MFM studies of
conventional superconducting films show that disordered GB regions can
act as vortex-pinning sites.\cite{ref6,ref7}

The broader high-value tail of the 5 mTorr force-imbalance distribution in \figref{5}{(e)} indicates greater spatial dispersion within the analyzed field of view. The narrower 7 mTorr distribution is compatible with a more uniform local landscape and does not imply weaker macroscopic pinning.

The interaction-energy maps in \figref{5}{(c,d)} show stronger local fluctuations in the 5 mTorr film and smoother variation in the 7 mTorr film. Because identical London parameters and analysis procedures were used, the contrast is interpreted only relatively.

Likewise, the 5 mTorr energy distribution in \figref{5}{(f)} is broader, whereas the 7 mTorr distribution lies closer to the ideal triangular-lattice reference. Together with the distance, coordination, \(g(r)\), and \(\lvert\psi_{6}\rvert\) metrics, this result supports reduced spatial heterogeneity after improving effective GB connectivity, without implying a defect-free long-range lattice.

\section{Conclusion}

We have demonstrated that sputtering-pressure-induced changes in
grain-boundary connectivity reorganize the vortex matter in NbTiN films.
Structural and transport measurements identify distinct intergrain
coupling and superconducting homogeneity in films deposited at 5 and 7
mTorr, while MFM directly reveals corresponding differences in vortex
ordering and spatial heterogeneity. Coordinate-based analysis further
shows that the local force imbalance and vortex--vortex interaction
energy deviate to different extents from those of an ideal triangular
lattice. These results highlight the dual role of grain boundaries as
vortex-pinning sites and as potential weak links, and establish
effective grain-boundary connectivity as a practical parameter for
controlling magnetic-field tolerance in NbTiN-based superconducting
devices.

\section{Data Availability Statement}

The data that support the findings of this study are available from the
corresponding author upon reasonable request.

\section{Supporting Information}

Additional film-growth and measurement details; resistive-transition
definitions and WHH extrapolation procedure; MFM acquisition,
vortex-coordinate extraction, and vortex-order analysis; London-model
definitions, assumptions, and finite-field-of-view limitations;
sensitivity of the local force-imbalance and interaction-energy
descriptors to \(T_{f}/T_{c}\) and \(\lambda(0)\); additional local
bond-orientational-order maps and short-range positional-correlation
metrics; and experimental and model-parameter tables (PDF).

\section{Acknowledgments}

The authors acknowledge Prof. Fang-Ting Lin's group at Shanghai Normal
University for providing the NbTiN thin-film samples used in this work.
This work was supported by the National Natural Science Foundation of
China (Grant No. 12174242) and the Science and Technology Commission of
Shanghai Municipality (Grant No. 24CL2901702). J.-Y. G. also
acknowledges support from the Program for Professor of Special
Appointment (Eastern Scholar) at Shanghai Institutions of Higher
Learning.

\FloatBarrier

\end{document}


\maketitle

\section{Additional Experimental Details}

The main text describes the overall film-growth and characterization procedures. Table S1 lists the additional conditions used for the direct comparison of the 5 and 7 mTorr films.

\begin{table}[htbp]
\centering
\caption{Additional film-growth and measurement details.}
\label{tab:S1}
\small
\begin{tabularx}{\textwidth}{>{\raggedright\arraybackslash}p{0.24\textwidth}XX}
\toprule
\textbf{Item} & \textbf{5 mTorr film} & \textbf{7 mTorr film} \\
\midrule
Substrate & Si(100), $10\times10\times0.5$ mm$^3$ & Same \\
Substrate cleaning & Acetone/ethanol/deionized-water ultrasonication; in situ Ar plasma cleaning for 300 s & Same \\
Base pressure & $5\times10^{-7}$ Torr & Same \\
Sputtering gas & Ar/N$_2$ = 81/9 sccm (9:1) & Same \\
Total sputtering pressure & 5 mTorr & 7 mTorr \\
Target composition & Nb:Ti = 70:30 wt\% & Same \\
Substrate temperature & Room temperature; no intentional heating & Same \\
Target--substrate distance & 70 mm & Same \\
Film thickness & Approximately 160 nm; stylus profilometry & Approximately 160 nm; stylus profilometry \\
XRD measurement & D2 diffractometer; Cu K$\alpha$; $\theta$--2$\theta$ scan; $2\theta=20$--80$^\circ$ & Same \\
Transport measurement & Four-probe resistance; 50 $\mu$A excitation current & Same \\
Magnetic sample dimensions & $2.31\times2.43$ mm$^2$ & $2.70\times1.84$ mm$^2$ \\
Magnetic background correction & A reference $M$--$H$ loop measured at 14 K under the same sweep protocol was subtracted from the lower-temperature loops & Same \\
\bottomrule
\end{tabularx}
\end{table}

\section{Analysis Criteria and MFM Procedure}

\subsection{Resistive-Transition and WHH Criteria}

For each $R$--$T$ curve, $R_n$ was determined by averaging the smooth normal-state resistance immediately above the superconducting transition. The onset temperature $T_{ci}$ was defined by $R/R_n$ = 0.90. For the 5 mTorr film, $T_{cj}$ marks the lower-temperature shoulder graphically and is used only as a qualitative indicator of intergranular coupling; it was not assigned a fixed resistance criterion and was not used in the WHH extrapolation. $R$--$T$ curves were measured from 2 to 18 K under fixed magnetic fields from 0 to 9 T during the field-increasing sequence. $\mu_0H_{c2}(0)$ was estimated from the near-$T_c$ slope using the dirty-limit orbital WHH relation and therefore represents an extrapolated, rather than directly measured, zero-temperature value.

\subsection{MFM Acquisition and Coordinate Analysis}

MFM images were acquired using an attocube attoDRY2100 low-temperature microscope and BudgetSensors Multi75M-G hard-magnetic MFM probes with a Co-alloy coating. After application of a field perpendicular to the film, the samples were cooled from above $T_c$ to 1.7 K. Magnetic contrast was acquired from the phase channel in noncontact lift mode. Vortex positions were identified from discrete contrast extrema and checked by visual inspection. The same imaging and coordinate-analysis procedure was applied to both films.

Delaunay triangulation and short-range positional and sixfold bond-orientational metrics were calculated from the extracted vortex coordinates using the same analysis settings for the two films.\cite{ref30,ref32}

\section{London-Model Assumptions and Sensitivity}

\subsection{Scope and Representative Parameters}

The interaction-energy and force-imbalance descriptors were recalculated from the measured vortex coordinates using the London vortex-line expressions given in the main text. The penetration depth at the effective freezing temperature was estimated using $\lambda(T_f)=\lambda(0)/\sqrt{1-(T_f/T_c)^4}$. The baseline values $\lambda(0)$ = 310 nm and $T_f/T_c$ = 0.92 are literature-guided representative calculation inputs, not independently measured material constants of either film. Published NbTiN penetration depths vary with film thickness and disorder, underscoring the need for the parameter sweep below.\cite{ref17,ref43}

Because a Pearl interaction is long-ranged, a finite MFM field of view does not contain all vortices contributing to the interaction, and the missing contribution is position dependent near an image boundary.\cite{ref36,ref37,ref38} The London calculation is therefore used only as a common local, semi-quantitative estimator for film-to-film comparison. It is not interpreted as an absolute microscopic pinning energy or as a direct measurement of pinning force.

\begin{table}[htbp]
\centering
\caption{Parameters used in the London-based sensitivity analysis at 80 Oe.}
\label{tab:S2}
\small
\begin{tabularx}{\textwidth}{>{\raggedright\arraybackslash}p{0.23\textwidth}>{\raggedright\arraybackslash}p{0.15\textwidth}XX}
\toprule
\textbf{Quantity} & \textbf{Baseline} & \textbf{Values used in sensitivity check} & \textbf{Role} \\
\midrule
$\lambda(0)$ & 310 nm & 280, 300, and 310 nm & Representative model input \\
$T_f/T_c$ & 0.92 & 0.90, 0.92, 0.95, and 0.98 & Representative model input \\
$\lambda(T_f)$ for $\lambda(0)=310$ nm & 582 nm & 529, 582, 720, and 1113 nm & Calculated from the two-fluid expression \\
\bottomrule
\end{tabularx}
\end{table}

\subsection{Sensitivity to the Effective Freezing Ratio}

At fixed $\lambda$(0) = 310 nm, varying $T_f$/$T_c$ from 0.90 to 0.98 changes $\lambda$($T_f$) and, consequently, the absolute scales of the calculated energy and force-imbalance descriptors (Figure~\ref{fig:S1}). These scale changes are expected from the London kernel. The same calculation is applied to both films at every parameter value.

\begin{center}
  \includegraphics[width=0.79\textwidth]{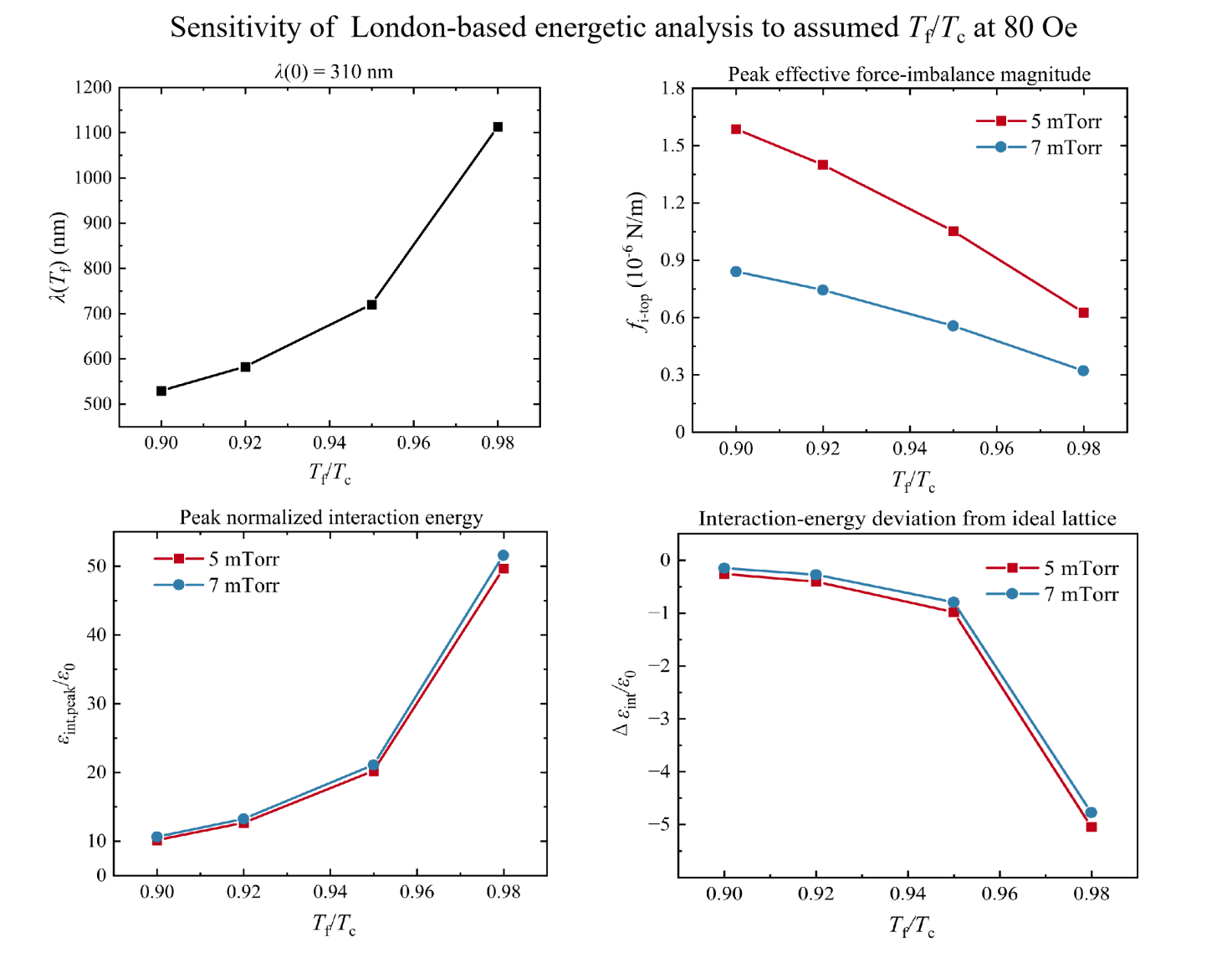}
  \captionof{figure}{Sensitivity of the London-based analysis at 80 Oe to the assumed $T_f/T_c$ at fixed $\lambda(0)$ = 310 nm. (a) Calculated $\lambda(T_f)$. (b) Peak effective force-imbalance magnitude. (c) Peak interaction energy normalized by $\varepsilon_0$. (d) Energy difference relative to the ideal triangular-lattice reference. The absolute values vary with the assumed freezing ratio; all quantities are used only for relative comparison under identical assumptions.}
  \label{fig:S1}
\end{center}

\subsection{Combined Sensitivity and Robust Comparison}

Figure~\ref{fig:S2} tests the conclusion drawn from the log-space interaction-energy distribution width, defined as $\sigma_{\ln}$ = std[ln($\varepsilon_{\mathrm{int}}$/$\varepsilon_0$)]. Across the investigated $T_f$/$T_c$ values, this width remains larger for the 5 mTorr film than for the 7 mTorr film. Varying $\lambda$(0) changes the absolute force-imbalance and energy offsets but does not reverse this contrast.

\begin{center}
  \includegraphics[width=0.90\textwidth]{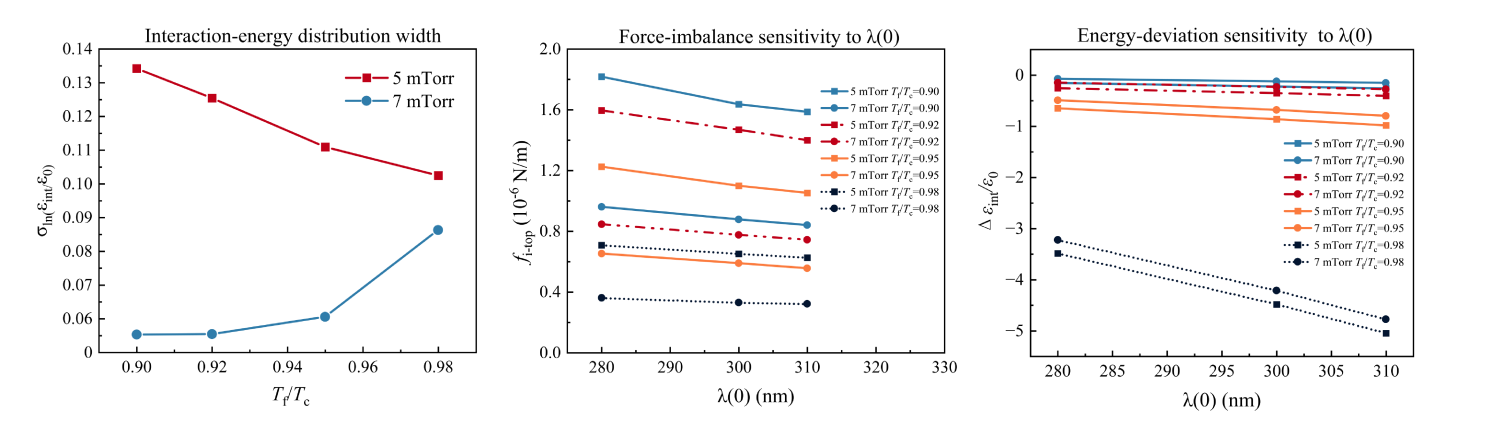}
  \captionof{figure}{Combined parameter-sensitivity check at 80 Oe. Left: log-space interaction-energy distribution width, $\sigma_{\ln}$ = std[ln($\varepsilon_{\mathrm{int}}$/$\varepsilon_0$)], as a function of $T_f/T_c$. Middle: peak effective force-imbalance magnitude as a function of $\lambda(0)$ for selected $T_f/T_c$ values. Right: interaction-energy deviation from the ideal-lattice reference. The width is used as a relative descriptor; the robust result is the broader distribution for the 5 mTorr film over the tested parameter range.}
  \label{fig:S2}
\end{center}

\section{Additional Real-Space Order Metrics}

The spatial maps of the local sixfold bond-orientational-order magnitude and the short-range pair-correlation metrics provide an additional view of the vortex configurations (Figure~\ref{fig:S3}). These panels supplement, rather than repeat, the distribution-based analysis in the main text.

\begin{center}
  \includegraphics[width=0.70\textwidth]{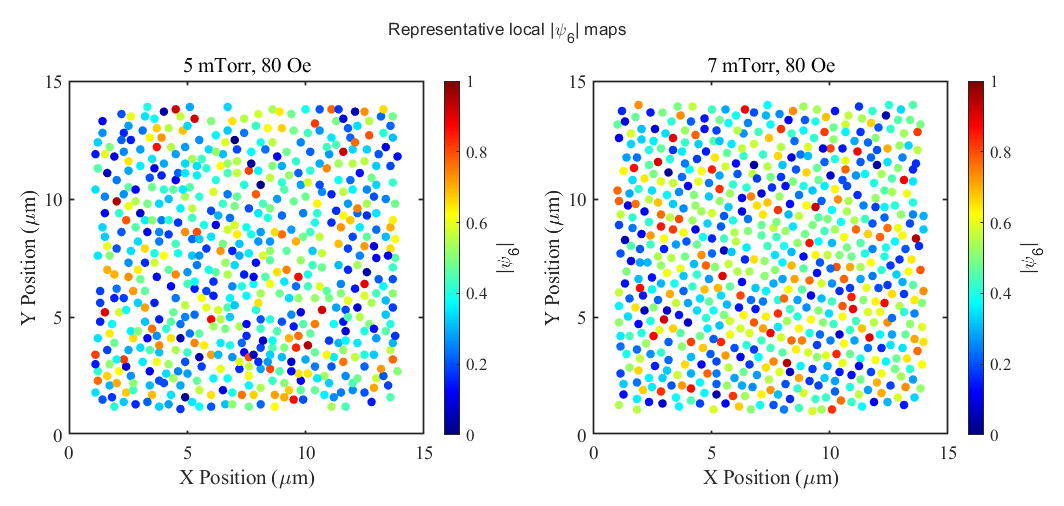}
\end{center}
\begin{center}
  \includegraphics[width=0.72\textwidth]{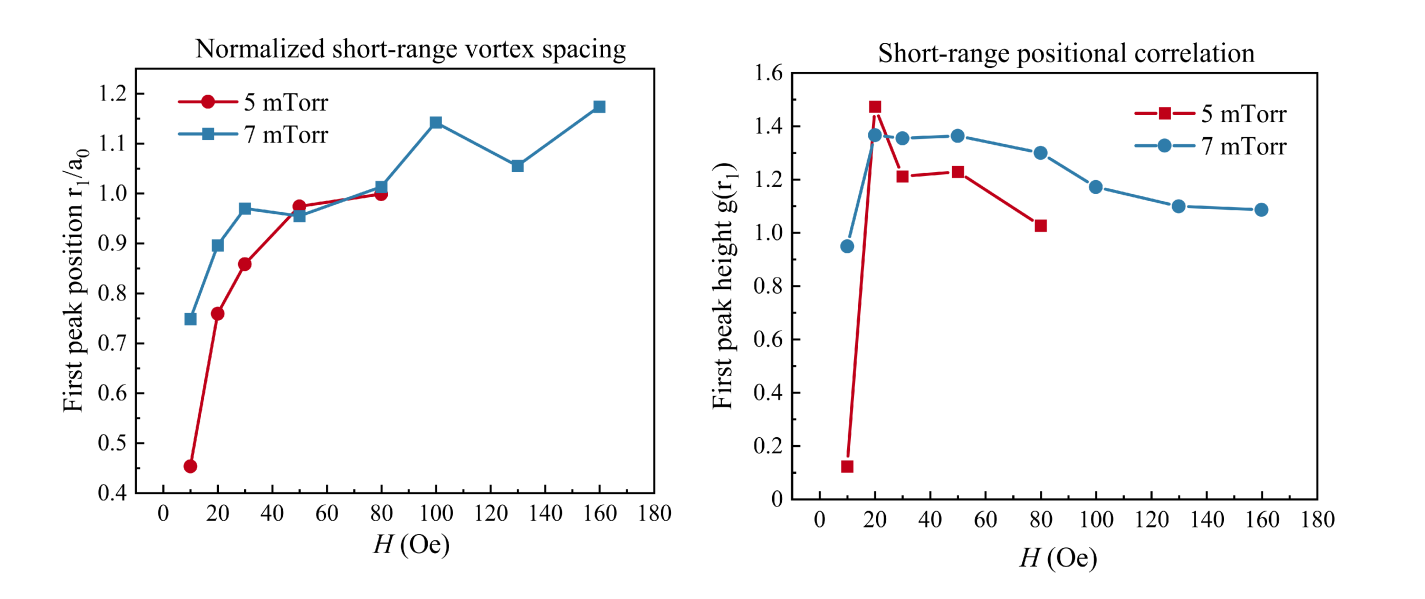}
  \captionof{figure}{Additional real-space order metrics. Top: maps of the local sixfold bond-orientational-order magnitude $|\psi_6|$ at 80 Oe for the 5 and 7 mTorr films. Bottom: field dependence of the position and height of the first peak of $g(r)$. The same coordinate-extraction and analysis procedure was used for both films.}
  \label{fig:S3}
\end{center}

Finite image boundaries can affect coordination statistics, pair correlations, and long-range interaction sums. Accordingly, the coordinate-derived energy and force quantities are interpreted only as local relative descriptors. No boundary-corrected Pearl interaction or deposition-to-deposition reproducibility is claimed from the present data.